\documentclass[twocolumn,showpacs,preprintnumbers,amsmath,amssymb]{revtex4}
\begin{document}
\title{Zero Modes in Light-Front $\phi^4_{1+1}$}

\author{G. B. Pivovarov}
\email{gbpivo@ms2.inr.ac.ru}
\affiliation{ Institute for Nuclear Research,\\
Moscow, 117312 Russia}

\date{January 4, 2005}
 
\begin{abstract}
Within a scheme of light front quantization of $\phi^4_{1+1}$, it is demonstrated that 
dynamics of zero modes implies phase transition, and that the critical value of the 
coupling coincides with the one of the equal time quantization.
\end{abstract}


\maketitle

There is a discussion in the literature on how phase transitions are relized
within the light front quantization of field theories (a discussion of this issue
and relevant references see in \cite{Kim:2003ha}). It is demonstrated in this note
that dynamics of zero modes can be completely decoupled from the rest of the 
dynamics, and implies the presence of phase transitions. The zero modes are the
modes of zero light-like momentum. In our consideration,
we use a regularization under which the modes of zero momentum are 
replaced with modes of low momenta (soft modes). Momenta of soft modes
vanish when the regularization is removed.

The regularization we use was introduced in 
\cite{Kim:2003ha}. It is a regularization introduced on the
level of the Lagrangian. The regularization involves higher derivatives to regularize
the ultraviolet divergencies, and a small external tensor coupled to derivatives of
the field under quantization. The external tensor breakes Lorentz invariance, and is needed
to avoid infinities caused by infinite volume of the Lorentz group. 
In particular, regularized Lagrangian of $\phi^4_{1+1}$ is
\begin{equation}
\label{regularized}
{\cal L}_r = \frac{1}{2}g^{\mu\nu}\phi_{,\mu}\phi_{,\nu}+
              \frac{1}{2}\epsilon t^{\mu\nu}\phi_{,\mu}\phi_{,\nu}+
               \frac{1}{2M^2}(\square\phi)^2
              - V(\phi),
\end{equation}
where $\square\phi=g^{\mu\nu}\phi_{,\mu\nu}$, and $V(\phi)=m^2\phi^2/2+g\phi^4/4$. The regularization is
removed
in the limit $\epsilon\rightarrow 0, M\rightarrow\infty$. The
tensor $t^{\mu\nu}$ is a symmetric tensor whose nonzero components in
a Lorentz frame are $t^{++}=1, t^{--}=-1$. The role of the term with
the $t$-tensor is to break Lorentz invariance. The signs of the
components of this tensor are chosen to have a nonnegative light-front
Hamiltonian (see \cite{Kim:2003ha}).

Standard quantization can be applied to the above Lagrangian, if 
$x^+\equiv(x^0+x^1)/\sqrt 2$ is taken as dynamical time, and the initial conditions
for the field are set on a light front at fixed $x^+$.

This quantization leads to a Hamiltonian, the operator, representing the $P_+$
compnent of the total momentum of the system. As explained in \cite{Kim:2003ha},
it is convenient to express the Hamiltonian in terms of the creation annihilation 
operators related to the field operator as follows:
\begin{equation}
\label{expansion}
\phi(x)=\int\frac{dk}{\sqrt{4\pi\omega_{lf}(k)}}
                       \left[a(k)e^{-ikx}+a^\dagger(k)e^{ikx}\right].
\end{equation}
Here $x$ is a shorthand for $x^-\equiv (x^0-x^1)/\sqrt 2$, and 
$\omega_{lf}(k)=\sqrt{k^2+(m^2+\epsilon k^2)(\epsilon + 4 k^2/M^2)}$.

In terms of the above creation annihilation operators, the Hamiltonian is
\begin{equation}
\label{hamiltonian}
H=\int dk\; \nu(k)\; a^\dagger(k)a(k)+ \frac{g}{4}\int dx\; \phi^4(x),
\end{equation}
where the energy of excitation is
\begin{equation}
\label{nu}
\nu(k)=\frac{\omega_{lf}(k)-k}{\epsilon+4k^2/M^2}.
\end{equation}

In the Hamiltonian, the normal product of fields is implied in $\phi^4$. 
Perturbatively, this Hamiltonian leads to Feynman diagrams that have finite 
limits at the regularization removed. These limits coincide with the expressions
obtained within the standard light front quantization, with the integrations
in the longitudinal momenta restricted to positive values.

The purpose of this note is to consider the limit of regularization removed
nonperturbatively.

Two observations will lead our consideration. First, gaussian effective
potential related to Hamiltonian (\ref{hamiltonian}) coincides in the
limit of regularization removed with the one computed for 
$\phi^4_{1+1}$ within equal time quantization, 
and implies the presence of phase transition \cite{Pivovarov}.
As noted in \cite{Pivovarov}, modes with momentum of order $\sqrt{\epsilon}$ are crucial
in the derivation. Second, if we neglect the modes of low momentum, 
and the states with light front energy going to infinity in the limit of regularization
removed, the Hamiltonian (\ref{hamiltonian}) is equivalent to the conventional
light front Hamiltonian (for the conventional light front Hamiltonian, 
see \cite{Brodsky:1997de}
and Eq. (\ref{lfhamiltonian}) below). 
Let us detail the second observation:
At fixed small or negative $k$, excitation energy $\nu(k)$ goes to infinity when the 
regularization is removed. Neglecting these infinite energy excitations, and replacing
$\omega_{lf}(k)$ with its limit ($\omega(k)\rightarrow k$), we arrive at the conventional
light front decomposition of the field in creation annihilation operators, and the
conventional light front Hamiltonian:
\begin{equation}
\label{lfexpansion}
\phi_{lf}(x)=\int_0^\infty\frac{dk}{\sqrt{4\pi k}}
                       \left[a(k)e^{-ikx}+a^\dagger(k)e^{ikx}\right],
\end{equation}
\begin{equation}\:
\label{lfhamiltonian}
H_{lf}=\int_0^\infty dk\; \frac{m^2}{2k}\; a^\dagger(k)a(k)+ \frac{g}{4}\int dx\; \phi_{lf}^4(x).
\end{equation}
In the last formula, we have taken into account that, at fixed positive $k$, 
$\nu(k)\rightarrow m^2/2k$ in the limit of regularization removed.

We conclude that Hamiltonian (\ref{hamiltonian}) can be decomposed as follows: 
\begin{equation}
\label{decomposition}
H=H_s+H_{int}+H_{lf},
\end{equation}
where $H_s$ is the contribution of the soft modes with momentum of order $\sqrt\epsilon$, and 
$H_{int}$ is the contribution to the Hamiltonian describing the interaction between 
the soft and conventional modes. Both $H_s$ and $H_{int}$ are negligible within perturbation
theory, but are important for understanding of the phase transition. 

To derive $H_s$ and $H_{int}$, we split the momentum axis into 
two domains, soft and normal. The soft momenta are restricted in magnitude with a cut $\Lambda_c$
that scales to zero as $\epsilon\rightarrow 0$, but slower than $\sqrt\epsilon$. In this way, the normal
modes occupy the whole momentum space at regularization removed, and, at the same time, the soft modes 
always contain momenta of order $\sqrt\epsilon$. For the soft modes, we rescale the momenta dividing them
by $\sqrt\epsilon$. Technically, this can be done via replacing creation annihilation operators
as follows:
\begin{equation}
\label{soft}
a(k) = \frac{1}{\epsilon^{\frac{1}{4}}}b(\frac{k}{\sqrt\epsilon}),\ 
a^\dagger(k) = \frac{1}{\epsilon^{\frac{1}{4}}}b^\dagger(\frac{k}{\sqrt\epsilon}).
\end{equation}
Notice that the scale of $b$, $b^\dagger$ is such that they obey the canonical commutation
relations, as $a$, $a^\dagger$ do (explicitly, $[b(k),b^\dagger(p)]=\delta(k-p)$). Our prescription
for studying the limit $\epsilon\rightarrow 0$ is to replace $a$-operators with $b$-operators each time
the momentum of $a$-operator is lower than $\Lambda_c$. In doing this replacement, the following object 
is of use:
\begin{equation}
\label{softfield}
\Phi_s(x)\equiv\int\;\frac{dk}{\sqrt{4\pi\omega(k)}}\:[b(k)e^{-ikx}+b^\dagger(k)e^{ikx}],
\end{equation}
where $\omega(k)=\sqrt{k^2+m^2}$. Notice, that this field is decomposed into $b$-operators
in the way taking place in equal time quantization. The last prerequisite we need for deriving $H_s$ and
$H_{int}$ is the relation between a piece of the field involving the soft modes,
\begin{equation}
\label{softlf} 
\phi_s(x)
\equiv\int_{-\Lambda_c}^{\Lambda_c}\;\frac{dk}{\sqrt{4\pi\omega_{lf}(k)}}\:[a(k)e^{-ikx}+a^\dagger(k)e^{ikx}],
\end{equation}
and $\Phi_s(x)$ of Eq. (\ref{softfield}). It is easy to establish in the limit $\epsilon\rightarrow 0$
the following relation:
\begin{equation}
\label{relation}
\phi_s(x)=\Phi_s(\sqrt\epsilon x).
\end{equation}
(The relations $\omega_{lf}(\epsilon k)\rightarrow\sqrt\epsilon\:\omega(k)$, 
$\Lambda_c/\sqrt\epsilon\rightarrow\infty$ should be used.)

Straightforward consideration leads to the following expressions for $H_s$ and
$H_{int}$:
\begin{equation}
\label{hsoft}
H_s=\frac{1}{\sqrt\epsilon}[H_{et}-P_{et}],
\end{equation}
\begin{equation}
\label{hint}
H_{int}=g\Phi_s(0)\int\:dx\;\phi_{lf}^3(x)+\frac{3g}{2}\Phi_s^2(0)\int\:dx\;\phi_{lf}^2(x).
\end{equation}
Here $H_{et}$ and $P_{et}$ are respectively the Hamiltonian and the momentum of
$\phi^4_{1+1}$ in equal time quantization:
\begin{equation}
\label{eth}
H_{et}=\int\;dk\;\omega(k)\:b^\dagger(k)b(k)+\frac{g}{4}\int\;dx\;\Phi_s^4(x),
\end{equation}
\begin{equation}
P_{et}=\int\;dk\;k\:b^\dagger(k)b(k).
\end{equation}
(The relation $\nu(\sqrt\epsilon k)\rightarrow(\omega(k)-k)/\sqrt\epsilon$ should be used in the 
derivation of Eq. (\ref{hsoft}), where $\nu(k)$ is defined in Eq. (\ref{nu}).)

We conclude that Hamiltonian (\ref{hamiltonian}) contains a piece that scales as
$1/\sqrt\epsilon$. Thus, this piece can give a finite contribution 
to the energy, if $H_{et}-P_{et}$ is of order $\sqrt\epsilon$, or zero. Fortunately, we know
that $H_{et}$ defines a theory with a mass gap (if the coupling does not
equal the critical value) \cite{Glimm:ng}. Thus $H_{et}-P_{et}$ can be of 
order $\sqrt\epsilon$ only if $P_{et}$ is of order $1/\sqrt\epsilon$, which is impossible, because
of the cutoff $\Lambda_c$. We conclude that the only way to keep the energy finite is to restrict the
state space of the soft modes to the vacum vector of $H_{et}$, in which case the contribution
proportional to $H_{et}-P_{et}$ vanishes.

After this conclusion, we see that $\Phi_s(0)$ and $\Phi_s(0)^2$ in Eq. (\ref{hint}) should be replaced
with their vacuum expectations. Ultimately, after the regularization is removed,
\begin{equation}
\label{ultimatum}
H=H_{lf}+
g\Phi\int\:dx\;\phi_{lf}^3(x)+\frac{3g}{2}\Phi^2\int\:dx\;\phi_{lf}^2(x),
\end{equation}
where $\Phi$ and $\Phi^2$ are the vacuum expectations of the field operator and field operator
squared (with normal product assumed) of $\phi^4_{1+1}$. At small coupling, both vacuum expectations
are zero, and $H=H_{lf}$. As we know from the properties of $H_{et}$, there is a critical value 
of the coupling beyond which the expectations are nonzero.

Let us summarise our findings. Dynamics of the soft modes is identical to the dynamics of 
the compete theory. It is decoupled from the dynamics of the normal modes. The only 
trace of the presence of the soft modes in the reguralized theory 
that survives the removal of the regularization is the phase transition. At coupling exceeding
the critical value, new terms should be introduced into the light front Hamiltonian to
account for nonzero vacuum expectations of the soft modes. 

This work was supported in part by RFBR grant no. 03-02-17047.


\end{document}